\input{aipcheck}

\documentclass[
    ,final            
  ]
  {aipproc}

\layoutstyle{6x9}

\begin{document}

\title{Search for the coolest white dwarfs in the Galaxy}

\classification{97.20}
\keywords      {(stars:) white dwarfs }

\author{S. Catal\'an}{
  address={Centre for Astrophysics Research, University of Hertfordshire, Hatfield, AL10 9AB, UK},
}

\author{R. Napiwotzki}{
  address={Centre for Astrophysics Research, University of Hertfordshire, Hatfield, AL10 9AB, UK},
}

\author{S. Hodgkin}{
  address={Institute of Astronomy, Madingley Road, Cambridge, CB3 0HA, UK},
}

\author{D. Pinfield}{
  address={Centre for Astrophysics Research, University of Hertfordshire, Hatfield, AL10 9AB, UK},
}

\author{D. Cristobal Hornillos}{
  address={Centro de Estudios de Fisica del Cosmos de Aragon (ceFca), E-44001, Teruel, Spain},
}

\begin{abstract}
A number of so-called ultra-cool white dwarfs have been detected in 
different surveys so far. However, based on anecdotal evidence it is 
believed that most or all of these ultra-cool white dwarfs are low-mass 
products of binary evolution and thus not representative for the oldest 
white dwarfs. Their low mass causes relatively high luminosity making 
them the first cool white dwarfs detected in relatively shallow surveys. 
Deeper observations are needed for the oldest, high mass white dwarfs 
with the longest cooling times. We report results of an ongoing project 
that combines deep IR and optical data. This combination plus proper 
motion information will allow an unambiguous identification of very cool 
white dwarfs, since the spectral energy distributions are very different 
from other types of stellar objects. The atmospheric parameters that can 
be derived from the spectral energy distributions together with the 
proper motions inferred from the IR data can be used to construct the 
white dwarf luminosity functions for the thick disc and halo 
populations. From these we will be able to test the early star formation 
history and initial mass function of the first stellar populations.
\end{abstract}

\maketitle


\section{Introduction}

Observations of high-redshift galaxies provide a direct, if blurred, 
view of the earliest phase of star formation in the Universe, with the 
initial starbursts having a big impact on further galaxy evolution. Only 
limited information is available from unresolved populations because 
observed spectral energy distributions (SEDs) are completely dominated 
by stars in a narrow mass interval close to the turn-off. Hence, 
measuring star formation rates requires the adoption of an initial mass 
function (IMF), with some authors favouring top-heavy prescriptions 
(e.g.~\cite{bau05}). A direct test is possible locally in our galaxy by 
studying thick disc and halo stars, which were formed at the same age as 
the starbursts observed in high-redshift galaxies.

Although all pop.~II stars more massive than the Sun evolved away from 
the main sequence long ago, the early IMF can be 
reconstructed from the luminosity function (LF) of the relic white dwarf (WD)
population. A thin disc WD LF has been 
measured based on SDSS \cite{har06}. Thin disc LFs 
are a convolution of the IMF, star formation history, 
initial-final-mass relation... and are almost impossible to invert. The 
formation history of the halo and thick disc are probably complex, but 
observations of field stars and globular clusters indicate that most 
stars are very old formed over a short period of time. So, the 
interpretation of the thick disc/halo LF is much more straightforward.

Ref.~\cite{opp01} reported a very large population of WDs 
belonging to the galactic halo -- in line with a top heavy IMF. This 
result was disputed by \cite{rei01} and \cite{pau06}, but all these 
investigations are based on very local samples and large uncertainties 
remain, because of small number statistics. A particular problem is the 
lack of confirmed very cool pop. II WDs, which are most 
interesting, because they evolved from the most massive progenitor 
stars. Even the pop. I WD LF constructed by 
\cite{har06}, comprising of 6,000 WDs selected from the Sloan 
Digital Sky Survey (SDSS) contains only 35 cool WDs with absolute 
magnitudes $>15$\,mag, thus, deeper observations are needed to increase 
this number.

We participate in a project that is performing a search of thick disc 
and halo WDs in a very deep infrared survey (WTS) carried out in the 
$zyJHK$ filters with the UKIRT telescope -- co-PI'ed by D.~Pinfield. The 
main aim of this survey is the detection of planetary companions to 
low-mass stars from the transit light curves. WTS will take over the 
time span of five years many repeat observations in NIR bands with a 
total coverage of 6 sq. deg. (4 fields). The $J$ band data will be the 
deepest with a total of 25 hours spent on each field, reaching at least 
$J=25$\,mag ($5 \sigma$) by co-adding observations. The $zyHK$ data have 
a depth of 19--21, so these are useful for bright objects. Our proper 
motion (ppm, $\mu$) work using stacked WTS observations from different 
epochs indicates a final ppm accuracy of $\approx 1$$-$$2$\,mas/yr. 50\% 
of the WTS observations are completed and reduced so far. Observations 
of the four fields are expected to be completed by 2012. We have 
recently obtained optical data ($g'r'i'$ bands) of one of the WTS fields 
with the CAHA 3.5m telescope and the LAICA detector. We have granted 
time to observe the rest of the fields.

\section{Simulations}

\begin{figure}
  \includegraphics[height=.4\textheight]{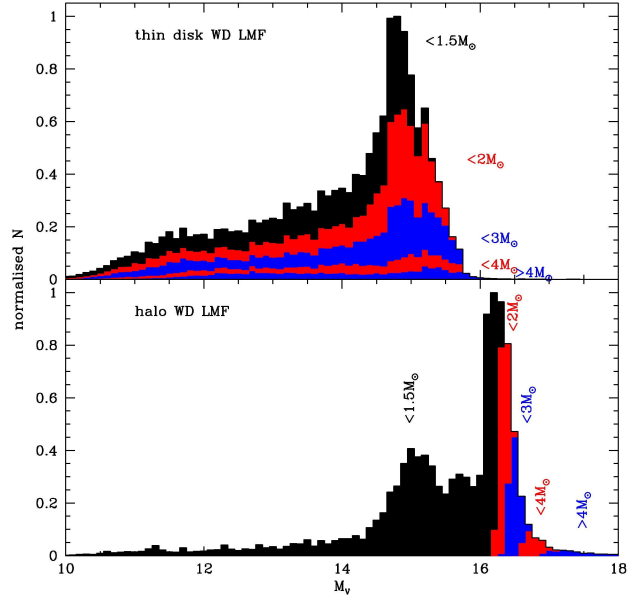}
  \caption{Simulated luminosity function of thin disc and halo WD population for a volume limited sample. The progenitor masses are colour coded for the intervals $M_{\rm 
prog}$$<$$1.5M_{\odot}$, $1.5M_{\odot}$$<$$M_{\rm prog}$$<$$2M_{\odot}$, 
$2M_{\odot}$$<$$M_{\rm prog}$$<$$3M_{\odot}$, $3M_{\odot}$$<$$M_{\rm prog}$$<$$4M_{\odot}$, 
$4M_{\odot}$$<$$M_{\rm prog}$. The halo WD luminosity function shows a huge pile-up 
in the coolest bins. These are WDs produced from higher mass progenitors within a 
few $10^8$ years after the first star formation bursts.}
\end{figure}

\begin{table}
\begin{tabular}{lccc}
\hline
  \tablehead{1}{r}{b}{IMF}
  & \tablehead{1}{r}{b}{$N(M_{\mathrm{prog}})>1.5M_\odot$}
  & \tablehead{1}{r}{b}{$N(M_{\mathrm{prog}})>2.0M_\odot$}
  & \tablehead{1}{r}{b}{$N(M_{\mathrm{prog}})>4.0M_\odot$}   \\
\hline
Salpeter & 27 & 11 & 1 \\
Baugh & 135 & 94 & 26 \\ 
Kennicutt & 49 & 26 & 2 \\
\hline
\end{tabular}
\caption{Total number of WDs expected in the four WTS fields.}
\label{tab:a}
\end{table}

Ref.~\cite{nap09} constructed a model of the Galactic WD population, 
based on the model of Galactic structure by \cite{rob03}. 
These simulations consider the WD cooling sequences from 
\cite{blo95}, the stellar tracks of \cite{gir00} (Padova Group) to 
account for the WD progenitor lifetime and the initial-final 
mass relationship of \cite{wei00}. The Salpeter IMF is assumed, although 
the simulations can be run with other IMFs (\cite{bau05}, \cite{ken83}). 
The population identification is then based on the results of the 
kinematic study of \cite{pau06}, calibrated with the local sample 
\cite{hol08} and checked against the proper motion selected sample of 
cool WDs by \cite{opp01}. According to these simulations more 
than 1,500 WDs with ppm $>$ 10\,mas/yr will be detected in the 
WTS fields. This model assumes a standard IMF, but numbers could be much 
higher for a top heavy IMF. Moreover, this sample will contain 
$\approx$100 cool WDs with $M_V>15$\,mag ($T_{\mathrm{eff}} 
<$5,000 K), most of them thick disc and halo members, with $M_V$ fainter 
than the drop-off in the thin disc LF. In Fig.~1 we show these 
simulations for the thin disc and halo WD populations and for different 
ranges of progenitor masses. As expected, the halo WD LF shows a huge 
pile-up in the coolest bins corresponding to WDs produced from higher 
mass progenitors within a few $10^8$ years after the first star 
formation bursts. The situation is different for the thin disk LF, 
because of the continuous, still ongoing star formation. In Table 1 we 
show the number of WDs with progenitor masses $M_{\rm prog}$ 
above threshold masses expected in the four WTS fields. As expected, the 
predicted number of WDs with massive progenitors is much higher 
if the Baugh et al. IMF is adopted. With observations for the four WTS 
fields we will be in a position of ruling out or confirm IMFs predicting 
an overproduction of high mass stars down to the Kennicutt prescription. 

We also checked with the simulated Galactic WD population that 
ppm criteria allow a good distinction between thin disc WDs on 
the one hand and thick disc and halo WDs on the other hand, 
which will allow us to define a clean, but still well sized sample of 
WDs from the earliest Galactic populations. Distinction between 
halo and thick disc members will not be as clear cut, but the main aim 
of this project -- reconstruction of the early Universe IMF -- will 
nevertheless be straightforward, because it can be based on the relative 
numbers of WDs in absolute magnitude (or $T_{\mathrm{eff}}$) 
bins.

\section{Characterization of white dwarf candidates}

\begin{figure}
  \includegraphics[height=.35\textheight]{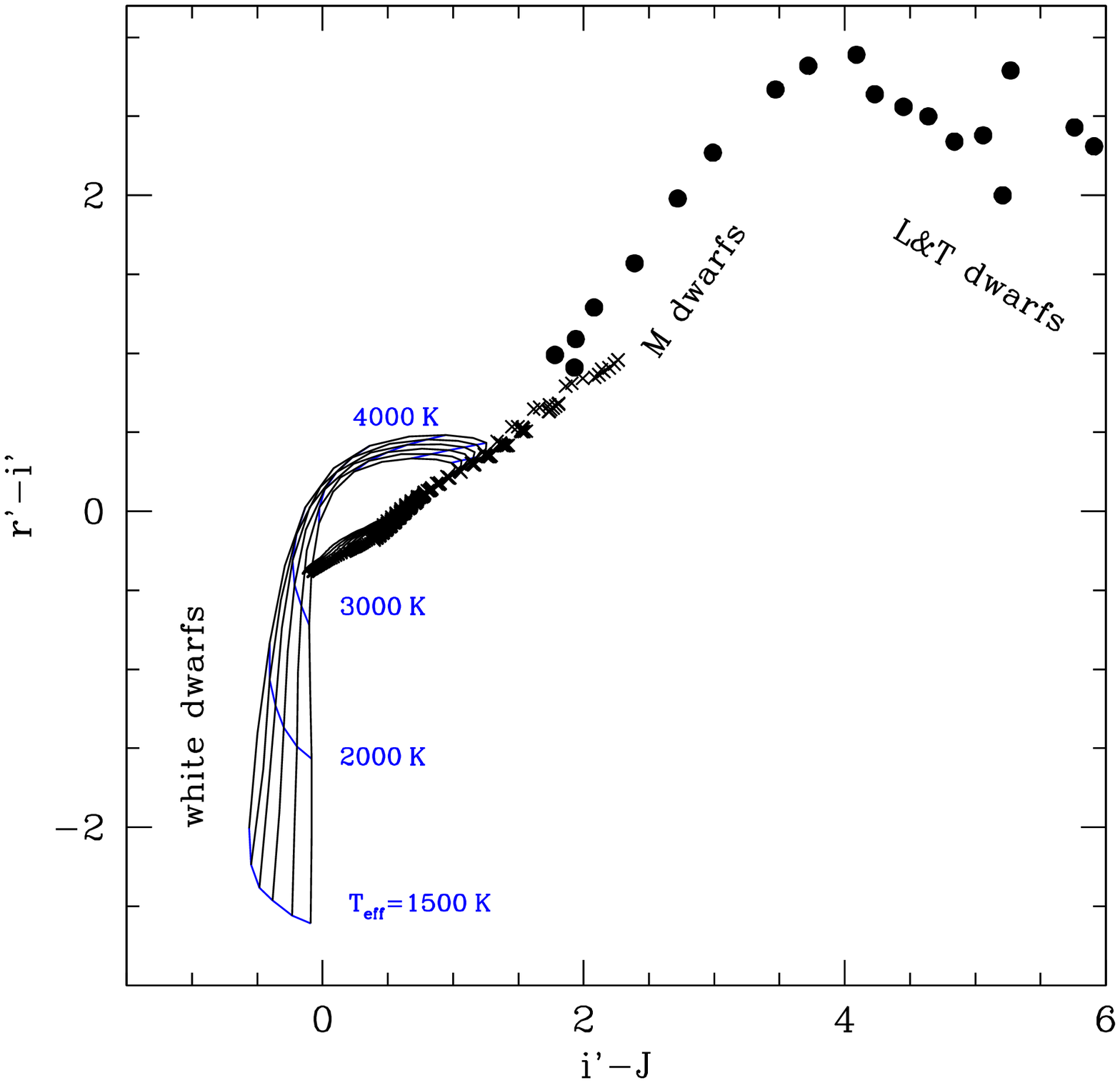}
  \includegraphics[height=.35\textheight]{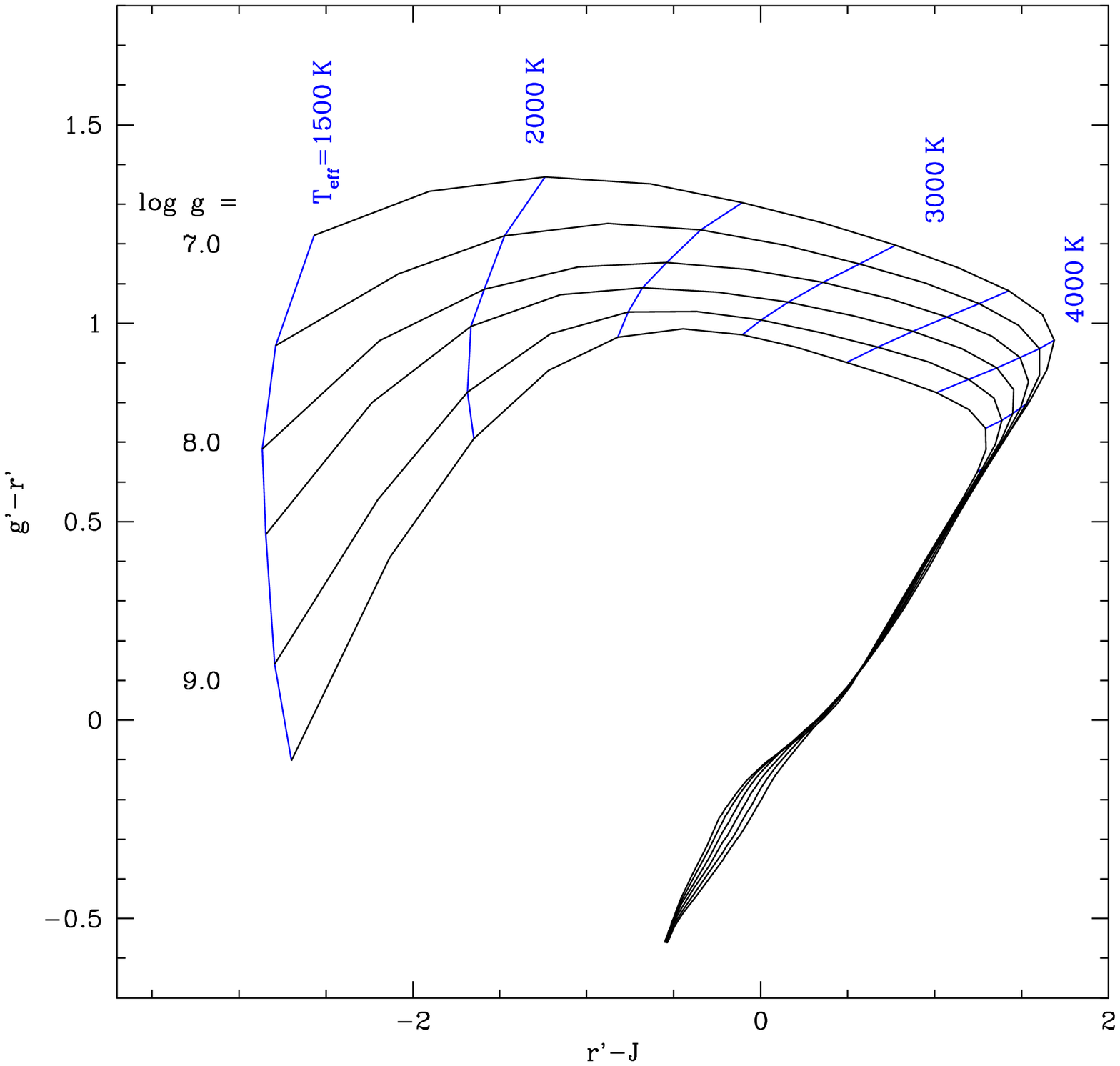}
  \caption{ {\it Left:} Colour--colour diagram for different cool objects. Filled circles are 
empirical colours of M dwarfs and brown dwarfs from \cite{haw02}. Crosses 
represent synthetic photometry from \cite{bes98}. The grid on the left 
corresponds to synthetic colours for WDs \cite{hol06} for different 
temperatures and gravities.  {\it Right:} Another colour--colour diagram that can 
be used to identify WDs showing good sensitivity in the cool domain.}
\end{figure}

A number of so-called ultra-cool WDs that show flux depression 
in the IR due to hydrogen molecule absorption have been detected so far 
\cite{har08}. However, it is believed that most or all of these 
ultra-cool WDs are low-mass products of binary evolution and 
thus not representative for the oldest WDs. Their low mass 
causes relatively high luminosity making them the first cool WDs detected 
in relatively shallow surveys. Deeper observations are 
needed for the oldest, high mass WDs with the longest cooling 
times. There are some available surveys, e.g. SDSS, that cover a larger 
area of the sky, but they are too shallow to detect these interesting 
objects. The combination of very deep IR observations with optical data 
will allow an unambiguous identification of cool WDs without 
further follow-up observations, since their spectral energy 
distributions are very different from other types of cool objects. In 
Fig.~2 (left) we show a colour--colour diagram for different cool 
objects. As it can be seen WDs can be easily discerned from other cool 
objects, and most importantly, the temperature and gravities of cool 
WDs can be obtained since it shows good sensitivity up to 
4,000 K, which is the domain in which we are mainly interested. In 
Fig.~2 (right) we show another colour--colour digaram showing good 
$T_{\mathrm{eff}}$ / $\log g$ sensitivity for the same domain. Hotter 
WDs are also needed to normalise the WD LF relative to lower mass progenitors. This can be achieved using 
a reduced proper motion diagram or other statistical tools \cite{nap08} 
even without direct measurements of $\log g$.


\begin{figure}
  \includegraphics[height=.35\textheight]{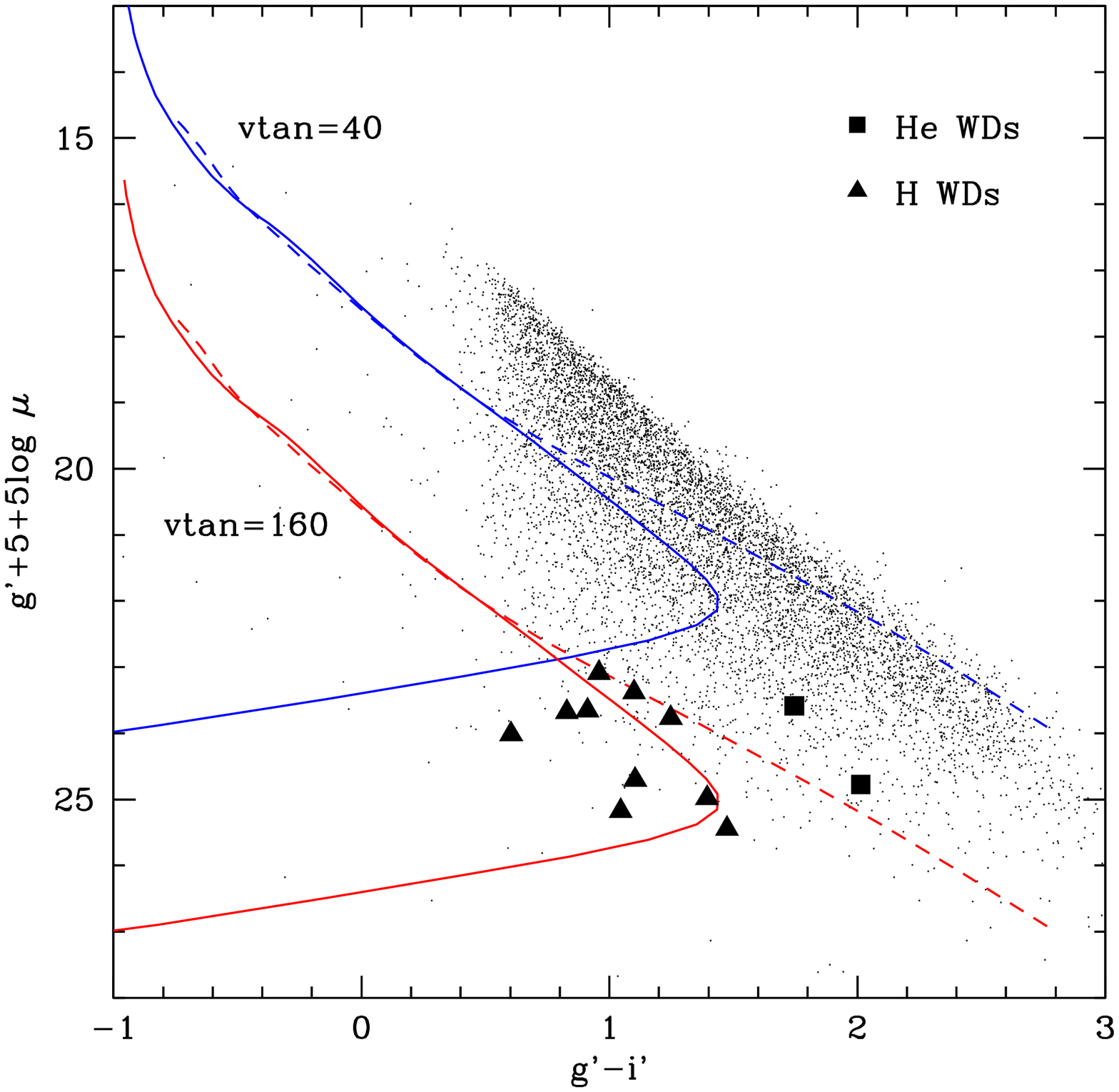}
  \includegraphics[height=.35\textheight]{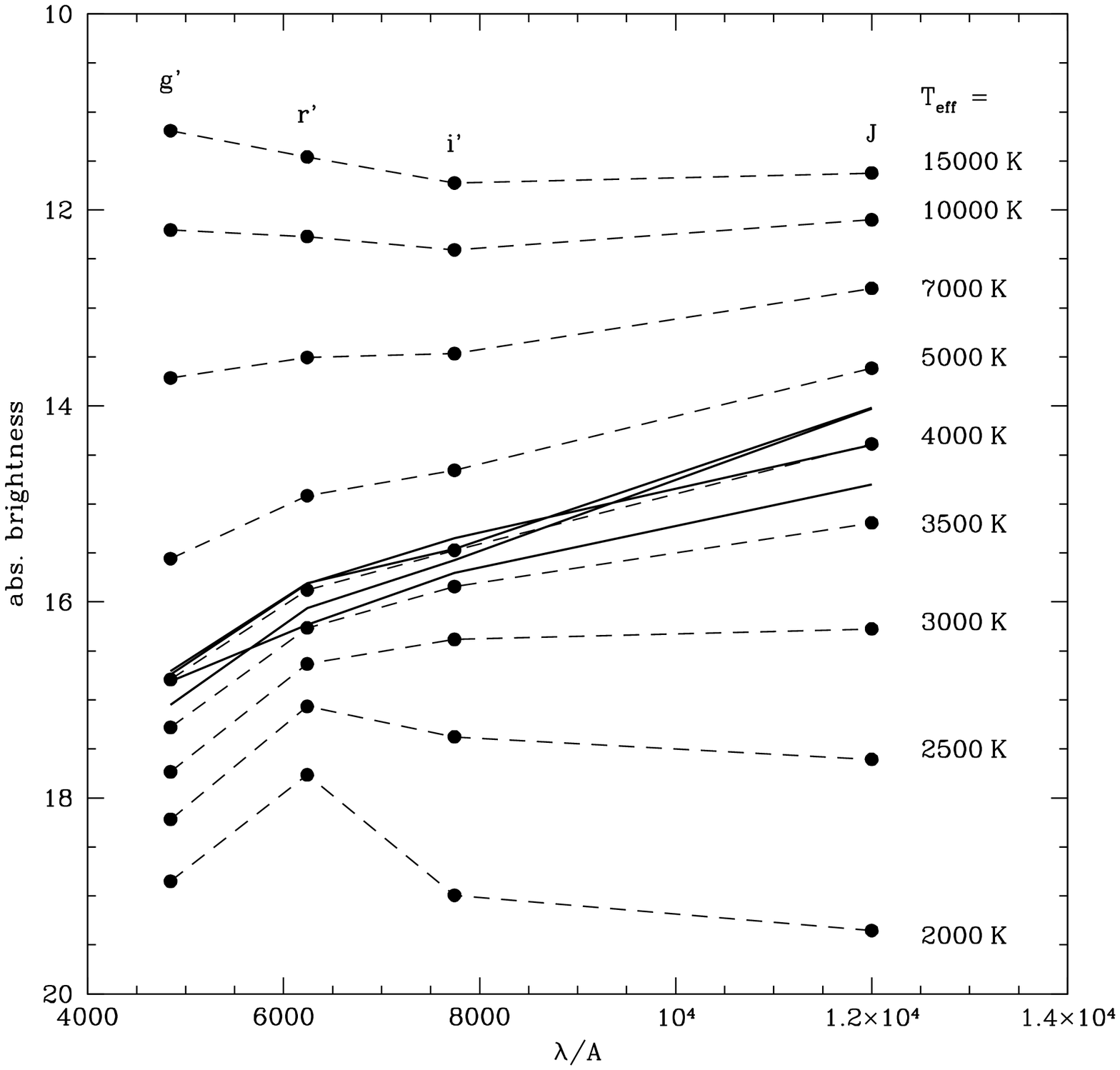}
  \caption{{\it Left:} Reduced proper motion diagram for the data obtained in the last observing run (May 2010). Triangles and squares correspond to our 12 high-ppm WD candidates. The solid lines correspond to the theoretical tracks for H composition for $v_{tan}=40$ km/s (top) and 160 km/s (bottom). The dashed lines correspond to He composition. Note that some of our WD candidates could have mixed compositions too. {\it Right:} Theoretical SEDs of WDs for different $T_{\mathrm{eff}}$ and $\log g = 8.0$. Solid lines correspond to some of our candidates.}
\end{figure}

From the combination of optical + IR photometry and the ppms ($\mu$) we 
have been able to identify thick disc/halo WD candidates. In Fig.~3 (left) we 
show the reduced proper motion ($H_g=g'+5+5\log \mu$) diagram that we 
obtained using our recent data. We have identified 12 candidates that 
meet the WD red. ppm criteria $H_g>15.136+2.727(g'-i')$ and have high 
ppm ($\mu>0.15$ arcsec/yr), indicating that they belong to the thick 
disc-halo populations. The triangles and squares correspond to Hydrogen 
and Helium WD candidates, respectively. The SEDs of WDs are very 
different from all faint objects showing significant ppm. In Fig.~3 
(right) we show the SED of WDs for different $T_{\mathrm{eff}}$ 
\cite{hol06}. The 2,000-3,000\,K models show the characteristic IR 
flux depression caused by hydrogen (pseudo)-molecules. The solid lines 
correspond to the magnitudes obtained for the coolest candidates and as 
it can be seen they match with the WD theoretical models. We have 
obtained preliminary $T_{\mathrm{eff}}$ and $\log g$ determinations for 
these 12 candidates from a fit to synthetic photometry, confirming that 
they are cool and have high $\log g$, consistent with single stellar 
evolution. From the atmospheric parameters, the WD masses can 
be derived using appropriate cooling sequences 
(e.g.~\cite{sal00},\cite{fon01}) and from these the progenitor masses by 
considering a initial-final mass relationship 
(e.g.~\cite{wei00},\cite{cat08}).

\section{Conclusions and Future work}

Thanks to the cool WDs detected in this project we will be able 
to reconstruct the early Universe IMF from a comparison of the LFs with 
the WD population simulations of \cite{nap09} until a self-consistent 
model is achieved. Some uncertainty is caused by the very poorly known 
initial-final mass relationship (IFMR) of pop.~II WDs, but even a modest variation of the IFMR will not 
have much impact on our project since our primary indicator of mass will 
be the WD temperature, and for the halo/thick disc LFs these are mainly 
determined by the progenitor lifetime, which is well known from stellar 
evolution models. However, 1) we will use the mass determinations of WDs 
in our sample to refine the IFMR, and 2) the resulting uncertainties are 
small compared to the effects claimed by \cite{bau05}.

In the near future our main objective will be to complete the optical 
imaging of the four WTS fields, then we will be in a position of ruling 
out or confirm IMFs predicting an overproduction of high mass stars down 
to the Kennicutt prescription. We will perform a spectroscopic follow-up 
of the coolest WD candidates to confirm their classification and study 
the atmospheric properties. The identification and characterization of 
cool WDs in this project will have an impact on other fields of 
astrophysics: Hydrogen opacities are rather uncertain at low 
temperatures. Theoretical work is going on for further improvement 
(e.g.~\cite{tre09}) but observational benchmarks are important. Note 
hydrogen have an impact on the IR flux of brown dwarfs as well, which 
are observed down to much lower temperatures than WDs \cite{all01}. We 
will obtain accurate observational SEDs of cool WDs that will give an 
unprecedented input to test the theoretical models.


\begin{theacknowledgments}
S.C. acknowledges financial support from the 
European Commission in the form of a Marie Curie Intra European Fellowship.
\end{theacknowledgments}



\bibliographystyle{aipproc}   





\end{document}